# BAR FORMATION, EVOLUTION AND DESTRUCTION

J. A. SELLWOOD

Department of Physics and Astronomy, Rutgers University, Piscataway, NJ 08855

## ABSTRACT

I review the various mechanisms for creating bars in rotating stellar disks, and conclude that the swing-amplified feed-back loop, which produces rapidly tumbling bars, remains the most probable. The bar continues to evolve after its formation in a number of ways; here I discuss what appears to be inevitable thickening normal to the plane, continued spiral activity in the outer disk and also underscore the increasingly important problem presented by angular momentum loss to the halo. Finally, I examine possible means, excluding interaction, by which bars in galaxies could be destroyed.

## 1. BAR FORMATION

### 1.1. *Global linear instability*

Bars in disk galaxies are widely believed to have formed through the well-known global dynamical instability first discovered in $N$-body simulations (Hohl & Hockney 1969, Miller, Prendergast & Quirk 1970). This instability afflicts many reasonable models of disk galaxies with significant rotational support and, somewhat to our embarrassment, the problem of the origin of bars in galaxies has been supplanted by the opposite problem of how can galaxies avoid forming a bar!

Toomre (1981) argued convincingly that the linear mode is an unstable standing wave between co-rotation and the galaxy centre, with the swing-amplifier causing leading waves incident on co-rotation to super-reflect into amplified trailing waves. Toomre claimed support for his mechanism from the existence of overtones, in addition to the fundamental bar mode, which appeared as slower modes in his linearised global analysis of cold disks. Quiet start $N$-body simulations, in which the amplitude of shot noise from the particles is greatly reduced by setting down particles in a non-random pattern, are in precise quantitative agreement with predictions from global linear theory and also have detectable overtones (Earn & Sellwood 1995).

Still more impressive support comes from the verification of Toomre's prediction that very high angular velocities near the galaxy centre should shut off the linear instability. This is because the feed-back cycle is interrupted by an inner Lindblad resonance (ILR) where trailing waves are absorbed – at least in linear theory. Quiet start simulations completely verified Toomre's unpublished global mode calculations (Sellwood 1989) and demonstrated that the linear bar-forming instability can be stabilized in an almost fully self-gravitating disk by the addition of a low mass, but dense, central bulge. This result stands as a counter-example to all claimed global stability criteria (e.g., Ostriker & Peebles 1973, Efstathiou, Lake & Negroponte 1982, Christodoulou, Tohline & Shlosman 1995, Christodoulou *et al.*, this meeting) which seem to give a reasonable guide only to systems with mild differential rotation.

---

[1] To appear in: IAU Colloquium **157** *Barred Galaxies*, edited by R. Buta, B. G. Elmegreen & D. A. Crocker

**Figure 1** The power spectrum (contours) of $m=2$ disturbances in a potential with a small core radius. The solid line indicates the circular frequency, $\Omega$, the dashed lines $\Omega \pm \kappa/2$. Above shows the early stages when the amplitudes of disturbances are small, and shows that an ILR is present. Below shows later in the same run when the bar is strong.

### 1.2. *Finite amplitude instability*

Toomre's mechanism for the global instability, may therefore seem to suggest that no galaxy having a steeply rising rotation curve ought to be barred. However, neither real galaxies nor $N$-body simulations blindly obey the laws of linear theory, and perturbations of finite amplitude can trigger the formation of a bar in models which linear theory would predict to be stable. In fact, merely shot noise caused by the random distribution of tens of thousands of particles can be swing-amplified sufficiently to destabilize a linearly stable disk and cause a bar to form (Sellwood 1989). Naturally, the larger the number of particles and the weaker the swing amplifier, the more difficult it becomes to trigger a bar in this way.

Results from a more recent example are illustrated in Figure 1, which shows the power spectrum from a noisy start 2-D simulation of a Fall & Efstathiou (1980) model with $V_{\rm m} = 0.8/\sqrt{GM_{\rm D}\alpha}$ and $ab = 0.1$. Even in the early stages, the pattern speed of the dominant $m = 2$ disturbance is always less than the maximum of the $\Omega - \kappa/2$ curve. Linear theory predicts that any such wave ought to damp at its ILR (Lynden-Bell & Kalnajs 1972, Mark 1947), but its amplitude is large enough that the resonance saturates; i.e., particles are trapped by the strongly non-axisymmetric potential into distorted orbits that are aligned. The result is once again a large-amplitude, rapidly-rotating bar that resembles in many respects those formed in models with more gently rising rotation curves.

The perturbation has to be strong enough to trap particles at the ILR at the first attempt; weaker spirals are damped at the resonance, heating the disk strongly and making it much harder for the non-linear behaviour to occur subsequently. In order to get the bar to form in this model, which employed 50K particles, a fine grid and short time step, I had to make the disk quite cool initially ($Q = 0.5$); similar models with $Q \gtrsim 1$ did not form a bar.

A distinctive feature of bars formed in this manner is that they appear to have an inner limiting radius inside which the density distribution is axisymmetric (Figure 1, lower panel), whereas the $m = 2$ component of density in bars formed by the linear instability declines more gradually towards zero at the centre. The absence of a coherent disturbance at the bar frequency (or any other) indicates there is no inner, perpendicular bar in this simulation, nor have I seen one in many other such simulations. The potential supports the inner perpendicular ($x_2$) orbit family, the generalization of the ILR for finite amplitude non-axisymmetric potentials, but apparently few particles in this purely stellar system librate around it.

Noise is not, of course, the only way to trigger a bar in a meta-stable disk galaxy. A tidal interaction with an external perturber could also achieve the same result, as has been demonstrated e.g., by Noguchi (1987). Thus it is by no means excluded that bars can form directly in galaxies where the central density is high enough to inhibit bars from forming by Toomre's small-amplitude mechanism.

### 1.3. *Attempts to form slow bars*

When the bar first forms through either of the above mechanisms, its semi-major axis corresponds pretty much to the co-rotation radius of the spiral-shaped disturbance which led to the bar. This is because particles are captured (rather suddenly) onto orbits aligned with the bar within this radius, while there can be little trapping beyond. Since the pattern speed does not change much during this non-linear saturation phase, the bar ends close to the major-axis Lagrange points $L_1$ and $L_2$; this is what is meant by a fast bar. Two groups have recently been attempting to form bars having much lower figure rotation rates, but both ideas face significant problems.

Combes & Elmegreen (1993) report cases in which the amplitude of the bar drops well inside the Lagrange points, though their models retain an aligned oval extension out to much larger radii (see their Figure 9). This happened in models started from a density distribution that was very abruptly truncated where the disk surface density was still high. I have been able to reproduce their final result, but I also noticed that edge-related instabilities were provoked which grew just about as fast as the usual bar mode. As these modes saturated, the torques associated with the edge modes appeared to interfere with the usual trapping of particles into the bar, preventing it from extending to anything like its usual radius. A comparison test with a similar model in which the outer truncation radius was moved a lot further out formed the usual fast bar. Thus their slow bar requires the disk to be sharply truncated before the surface density has declined much below its central value; it is hard to imagine how such disks could be formed in nature.

Polyachenko & Polyachenko (1994) argue that slow bars can be formed through Lynden-Bell's (1979) mechanism, which is the radial orbit instability in disk geometry. For this to happen, the disk must be sufficiently hot that the usual rapid bar instability is suppressed, i.e., one in which the rms radial velocity exceeds the mean orbital velocity over a large fraction of the disk. They find that slow bars form in simulations of such models in which the particles are confined to a plane. These authors are aware (e.g., their poster at this meeting) that the large radial velocities in their model will make it highly susceptible to the bending instability (e.g., Sellwood & Merritt 1994). Figure 2 shows the 3-D evolution of a fully self-gravitating disk model having a much smaller degree of pressure support than they require, which both puffs out of the plane and forms the usual rapid bar. If random motion were sufficiently increased to suppress the usual rapid bar mode, the distribution would quickly become so puffed up as to no longer resemble a disk. Interestingly, the poster paper by these authors reports that immersing their hot disk in a rigid halo appears to reduce substantially the thickness produced by the bending instability, which may yet save their idea. It will be important to understand why a halo seems to have this effect.

**Figure 2** The evolution of a reasonably hot disk without a halo. The thin disk both bends out of the plane, becoming much fatter in the centre, and forms the usual fast bar.

## 2. BAR THICKENING

Once a bar has formed in a disk, its interesting evolution has only just begun! Here I focus on four aspects of bar evolution which do not require external interference; Athanassoula (this meeting) discusses aspects of encounters with companions which affect bars.

Many recent simulations have confirmed the original result of Combes & Sanders (1981) that bars formed in thin disks will thicken into peanut-shaped objects. All bars formed in simulations having sufficient resolution have, without exception, thickened in this manner.

The thickening mechanism was most clearly demonstrated by Raha *et al.* (1991) as the saturation of an out-of-plane bending instability. This collective bending instability was first discussed in a local approximation by Toomre (1966) but Merritt & Sellwood (1994 [MS]) showed that the local approximation overestimates the stabilizing effect of gravity for large-scale modes. MS also noted that the instability requires a supporting response from orbits, which limits the thickness of a stellar system that could possibly be bending unstable; as for any harmonic oscillator, the frequency of forcing has to be below the natural frequency, in order to produce an in-phase response. Thus particles which experience a vertical driving frequency, through their horizontal motion across a bend, can support the bend only if the driving frequency is less than their natural vertical oscillation frequency. This is, of course, merely a necessary condition for instability, since gravity provides an additional stabilizing force, but MS showed that gravity is so weak for global bending modes that this minimal frequency condition is effectively the main stability condition.

In effect, a thin bar will puff up through bending instabilities until the density in the mid-plane drops to a low enough value that the natural vertical frequency for a large fraction of particles drops below the forcing frequency from this global bend. Effectively this same argument was expressed by Pfenniger & Friedli (1993), who emphasized that the 2:2:1 resonance for particles near the mid-plane seemed to be expelled from the bar as it puffed up. See also Merrifield's paper in these proceedings for another, highly simplified, argument that leads to the same thickness limit.

The principal orbit family supporting a thick bar is the $2:2:1_a$ family (in the notation of Sellwood & Wilkinson 1993). These are the usual $x_1$ orbits in the plane twisted anti-symmetrically about the mid-plane so that a particle following the periodic orbit goes both in and out radially and up and down vertically twice for each time around the bar centre (in a frame which rotates with the bar). Miller & Smith (1979) were in fact the first to note that large numbers of particles were librating around this orbit family in their rotating thick bar.

A thickness criterion based on ratios of oscillation frequencies is not easily translated into an observationally testable prediction. However, as bars are thought to be populated by orbits of greater eccentricity than those in the axisymmetric part of the disk, it is likely that all bars have a vertical thickness greater than that of the outer disk in the same galaxy. Most simulators, beginning with Combes & Sanders (1981), have speculated that this is the origin of "peanut bulges", and the idea is now finding some observational support (e.g., Merrifield, this meeting).

**Figure 3** The evolution (in 2-D) of an exponential disk immersed in a rigid halo. The core radius of the total potential is small enough to cause a short bar to form by time $\sim 45$, but the bar continues to grow as the evolution proceeds.

## 3. BAR-DISK INTERACTION

My objective for the second of my topics is merely to draw attention to some ancient work of mine (Sellwood 1981) which may have been forgotten. I found that spiral activity occurs for a protracted period in the outer disk when the initial bar ends at a radius well inside the outer edge of the disk.

A further example, which has an approximately exponential mass distribution in a logarithmic potential with a harmonic core, is shown in Figure 3. This noisy start model forms a short bar by time 90 through the usual linear instability (though a three-armed pattern dominated at time 30); the bar then grows very markedly in length through subsequent interaction with spirals in the outer disk. As it grows, the pattern speed also drops in such a way that the Lagrange points are always just outside the bar – as far as one can tell in this rapidly evolving potential.

Note that once again, spiral activity in the disk outside the bar is quite clearly not driven by the bar; in this example, the strongest spirals are often three armed. I emphasize this point since it reinforces the earlier result of Sellwood & Sparke (1988) that the outer spiral can have a quite different pattern speed from that of the bar.

The bar appears to grow by trapping additional particles which are ready to give up angular momentum near the inner ends of bi-symmetric spirals. The new particles added to the bar in this way still have too much angular momentum to sink deep into the bar, and are therefore added to the outer end of the bar. Even though the bar is slowing down, its angular momentum content is rising.

## 4. BAR-HALO INTERACTION

It has long been clear to Tremaine (unpublished) that a bar rotating within a halo of collisionless particles should experience a drag due to dynamical friction, but the first calculation of the magnitude of the drag was made by Weinberg (1985). His estimate from perturbation theory, supported by semi-restricted $N$-body simulations, indicated that a fierce torque would sap the angular momentum of the bar on the embarrassingly short time scale of a few rotation periods.

I had earlier reported (Sellwood 1980) some preliminary evidence that a halo does indeed experience a strong torque in a fully self-consistent simulation, albeit with very poor spatial resolution and not integrated for long. Most subsequent work has been limited in other ways; e.g., Little & Carlberg (1991) restricted their calculation to a plane, using a hot, but flat population of particles to represent the halo and Hernquist & Weinberg (1992) used a rigid, unresponsive bar rotating in a responsive halo. Again both studies found significant torques, but neither answered the central issue of how a real galaxy would respond to strong secular torques applied for a Hubble time.

As computing power and $N$-body algorithms improve, we have reached the point at which we can begin to simulate the long-term evolution fully self-consistent disk + halo models in three dimensions with adequate spatial resolution and large numbers of particles. Some preliminary calculations using grid methods are reported by Debattista & Sellwood (this meeting) while Athanassoula (also this meeting) has been using a direct-$N$ algorithm on a GRAPE device.

The first results confirm once again that the bar is very substantially braked by friction with the halo. In our simulation, we measure a distortion in the halo density distribution which lags the bar and gives rise to the torque. However, the torque drops to near zero before the bar stops rotating, at which point the halo distortion is roughly aligned with the bar, as it must be. Somewhat surprisingly, this happens before the halo has been torqued up sufficiently to co-rotate with the bar, and suggests that the resonances, which give rise to the out-of-phase response, may have saturated.

I find even more remarkable the fact that the bar could survive after suffering a loss of a large fraction of its angular momentum while its pattern speed drops by a factor of five! This typical bar is extremely robust, and does not support the suggestion (e.g., Kormendy 1982) that bar strength might decay over time.

Weinberg's prediction that a strong bar should slow down dramatically in a massive halo has been qualitatively confirmed. By the time the torque becomes negligible, which takes somewhat less than a Hubble time for a reasonable scaling of our model, the distance from the centre to $L_1$ is more than twice the semi-major axis of the bar.

We believe bars in galaxies rotate much more rapidly than this, though the evidence is not conclusive. In most cases, we associate dust lanes with shocks in the gas flow and then appeal to hydrodynamical simulations (Athanassoula 1992; Weiner, Sellwood & Williams this meeting) which require a fast bar to set up a similar flow pattern. Other indirect evidence comes from work on rings (Buta, Combes, this meeting). One much more direct estimate of a high pattern speed is reported by Kuijken & Merrifield (this meeting) for the SB0 galaxy NGC 936.

Weinberg suggested that this discrepancy between theory and observation may indicate that (1) bars are weak, (2) halos are not very massive or (3) that angular momentum is added to the bar from the outer disk at a rate sufficient to compensate for that lost to the halo. Two further possible escapes from this increasingly serious dilemma are that, (4) the halo may be locked into resonance with the bar, or (5) bars in galaxies, as opposed to simulations, do not survive for long.

None of these alternatives seems attractive, however. (1) Weak bars are hard to reconcile with the observed strongly non-circular motions in the stars and gas (e.g., many papers at this meeting). (2) The evidence for massive halos, in conventional Newtonian mechanics, is strong even for barred galaxies (Bosma, this meeting). (4) While most halo mass lies at large radii, much halo material outside the core should still couple to the bar through resonances; it is therefore hard to believe that all major resonances become saturated before the bar slows significantly. The fast bar in NGC 936 is problematic for both (3) and (5). The spirals in this SB0 galaxy are extremely weak and cannot be adding much mass to the bar. Finally, the absence of young stars and gas makes it unlikely that the bar is young, but this possibility cannot be entirely ruled out; one could argue that the vestigial spirals in this galaxy are the remnant of a slow instability that has just now saturated, and that the galaxy was tipped over the stability boundary just as the gas was used up or removed.

## 5. BAR DESTRUCTION

Bars in $N$-body simulations have been found to be extremely robust; they form readily and can survive a long-lasting secular drag (as just noted), or quite major surgery (Sparke & Sellwood 1987). There have, however, been a few suggestions of ways bars in galaxies could be destroyed.

One of the most obvious is through interaction between a barred galaxy and a dwarf companion, which is discussed at this meeting both by Athanassoula, and by Nishida & Wakamatsu. Of course, a merger with a larger galaxy will destroy not only the bar, but the disc also!

Raha *et al.* (1991) speculated that the bending instability could possibly be violent enough to destroy a bar completely. They noted that the bar was weakened more when the instability grew vigorously but neither they, nor anyone else to my knowledge, has constructed a model in which the instability was completely disruptive. On the contrary, the bar frequently recovers to its approximate original strength in the long-term evolution. This idea seems not to be a viable method of bar destruction, therefore.

Another idea does look promising, however, namely the build up of a large concentration of mass at the centre of the bar. This idea, which has been developed over a number of years (Hasan & Norman 1990, Hasan, Pfenniger & Norman 1993), has received more attention recently since the discovery of large concentrations of molecular gas near the centres of bars (see reviews by Kenney and by Turner at this meeting). At least three groups (Wada & Habe 1992; Friedli & Pfenniger 1991; Friedli & Benz 1993; Heller & Shlosman 1994) have taken up the daunting challenge of trying to simulate self-gravitating gas inflow in bars. The idea continues to be vigorously explored by a number of workers (e.g., Hasan, Sellwood & Norman 1993; Friedli 1994; Nishida & Wakamatsu 1995) but there is still much to be done.

A fully three-dimensional calculation of a model otherwise similar to that reported by Hasan *et al.* (1993) is shown in Figure 4. The disk in this model is a Kuz'min/Toomre disk of length scale $a$, the gravitational potential of which is supplemented by a rigid Plummer sphere of scale size $0.4a$ and containing 30% of the total mass, $M$. This rigid component can be thought of as representing a bulge. Adopting $M$ and $a$ as units of mass and length and setting $G = 1$, time is reckoned in units of $\sqrt{a^3/GM}$. Choosing $M = 10^{10} M_\odot$ and $a = 2$ kpc implies a time unit of 13 Myr.

The gravitational field of the 200K disk particles was determined on a three-dimensional cylindrical polar grid having 65 radial, 80 azimuthal and 225 vertical mesh points. Particles passing close to the central mass concentration require extremely short time-steps ($10^{-3}$ time units) but it is inefficient to use this step size for all particles; I therefore adopted a three-zone stepping procedure (Sellwood 1989) in which time steps were increased by factors of 10 and then 5, for particles progressively further from the centre.

The first moment, time 100, shown in Figure 4 illustrates the disk component once a bar had formed, thickened and settled. From this moment to time 140, I gradually reduced the core radius of one sixth of the bulge mass, i.e., 5% of the total mass, from $0.4a$ to $0.02a$. This procedure does not add any mass to the system and is intended to mimic the radial inflow of gas driven by the bar. The bar responds to this change by first increasing its pattern speed and becoming shorter and then disrupts very abruptly between times 130 and 140. The rapidity of the decay is illustrated in Figure 5, the amplitude of the $m = 2$ harmonic drops by a factor 10 in less than one bar rotation period, after which time no significant bar remains. Interestingly, the mass distribution in the bar region thickens up quite markedly as the bar dissolves, forming a spheroidal, bulge-like distribution.

**Figure 4** The later stages of evolution of a bar-unstable disk with a rigid bulge as 5% of the total mass is compressed into a dense object in the centre. The bar formed and thickened during the period before the first frame. The central density was increased gradually from time 100 to 140, after which it was held constant. The bar disrupts into a spheroidal shape between times 130 and 140.

**Figure 5** The amplitude of the bar as a function of time in the model shown in Figure 4. The oscillations are due to beats between the bar and weaker spirals in the outer disk.

The demise of the bar on the shortest possible time-scale seems to occur because particles cease to be trapped about the main $x_1$ orbit family as the potential changes. The growth of the central mass alters the shapes of the periodic orbits and particles which had been moving on regular orbits probably find themselves in stochastic regions. The widespread breakdown of invariant tori in this rotating tri-axial potential leads to a brief period of chaos in which the orbits of particles are bounded only by their, much rounder, energy surfaces. A new equilibrium is quickly reached when the potential becomes axisymmetric, for which phase space is likely to be regular. This interpretation accounts both for the abrupt destruction of the bar, and for the vertical thickening of the particle distribution.

A number of authors have offered a different interpretation of the bar dissolution. They suggest that particles are individually scattered from their bar-supporting orbits as they pass by the central mass, causing the bar to be eroded more gradually. This argument probably stems from that given by Gerhard & Binney (1985) for non-rotating tri-axial ellipsoids; many orbits in these objects are boxes, which take stars close to the centre where a steep density gradient might change the orbit drastically and destroy the tri-axiality. [Merritt & Fridman (1995) show that this does not, in fact, preclude tri-axiality.] The main ($x_1$) orbit family in a rapidly rotating bar, on the other hand, are loops which always avoid the centre. When we consider central masses small enough to cause only partial disruption of the bar (Hasan *et al.* 1993, Friedli 1994), many stars will be pursuing well behaved loop orbits determined by the combined potential of the bar and central mass. It could not therefore be argued that deflections accumulate as a star repeatedly passes the centre and the gradual erosion picture that the scattering argument conjures up is misleading. In effect, focusing on the scattering of a test particle by an isolated point mass neglects the existence of the rest of the bar.

The one simulation reported here is very preliminary; it is clear that 5% of the entire galaxy mass in a very dense centre is enough to destroy the bar. Hasan *et al.* (1993) find that smaller masses weaken the bar, and Friedli (1994) reports that 3% in a point-like mass is enough to destroy it. However, we cannot yet say how the critical mass varies with concentration, which we need to know before we can predict how close barred galaxies may have come to the point of destruction.

**Acknowledgments** This work was supported by NSF grant AST 9318617 and NASA Theory grant NAG 5-2803.